\documentclass[showpacs,preprintnumbers,prb,twocolumn,floatfix]{revtex4}
\usepackage[dvips,xdvi]{graphicx}
\usepackage{epsfig}
\usepackage{pstricks}
\usepackage{pst-node}
\usepackage{pst-coil}
\usepackage{amsmath}
\usepackage[section]{placeins}

\begin {document}
\title{Hybridized quadrupole-dipole exciton effects in $Cu_2O$ - Organic Heterostructure}
\author{Oleksiy Roslyak and Joseph L. Birman}
\affiliation{Physics Department, The City College, CUNY\\
Convent Ave. at 138 St, New York, N.Y. 10031, 
USA}
\date{\today}
\begin{abstract}
In the present work we discuss resonant hybridization of the $1S$ quadrupole Wannier-Mott exciton (WE) in a $Cu_2O$ quantum well with the Frenkel (FE) dipole exciton in an adjacent layer of organic DCM2:CA:PA. The coupling between excitons is due to interaction between the gradient of electric field induced by DCM2 Frenkel exciton and the quadrupole moment of the $1S$ transition in the cuprous oxide. The specific choice of the organic allows us to use the mechanism of 'solid state solvation' (Bulovic \cite{MADIGAN:2003}) to dynamically tune the WE and FE into resonance during time $\approx 3.3 \: ns$ (comparable with the big life time of the WE) of the 'slow' phase of the solvation. The quadrupole-dipole hybrid utilizes the big oscillator strength of the FE along with the big lifetime of the quadrupole exciton, unlike dipole-dipole hybrid exciton which utilizes big oscillator strength of the FE and big radius of the dipole allowed WE. Due to strong spatial dispersion and big mass of the quadrupole WE the hybridization is not masked by the kinetic energy or the radiative broadening. The lower branch of the hybrid dispersion exhibits a pronounced minimum and may be used in applications. Also we investigate and report noticeable change in the coupling due to a induced 'Stark effect' from the strong local electric field of the FE. We investigated the fine energy structure of the quantum well confined ortho and para excitons in cuprous oxide.           
\end{abstract}
\maketitle
\section{\label{intro} introduction}

Nanometer-sized organic and inorganic semiconductor structures have recently been attracting much attention. In these low-dimensional systems, there are pronounced quantum confinement effects on the electronic and optical properties. Synthesizing composite organic/inorganic semiconductors is of  major  importance not only in the development of novel nano-structure materials for electronics, optics and transport, but also for basic understanding of their size-dependent physical properties.  Recently, a new type of elementary state which can be generated by optical excitation was discussed by Agranovich \cite{AGRANOVICH:1998}. This is a hybrid exciton which can be obtained from the resonant mixing of Frenkel(FE) and Wannier-Mott(WE) excitons in organic-inorganic quantum wells by means of dipole-dipole interaction across the interface. Many properties of this hybrid were predicted. Other realizations for the hybrid have been proposed. Examples are hybrid excitons in an inorganic semi-conducting quantum dot covered by an organic layer\cite{ENGELMAN:1998} and quantum dot-dendrimer system \cite{HUONG:2000}. The energy of the hybrid exciton as well as the Green's function matrix elements for different quantum dot-dendrimer systems has been calculated.
\par
In the model of Agranovich et. al. the decisive role in implementing the formation of the hybrid state is played by the dipole-dipole coupling between semiconductor WE and organic FE. It is assumed that there is no direct wave function overlap between the excitons on each side of the well. The interaction energy takes the form of $P\left(\bf{r}\right) \cdot E\left(\bf{r}\right)$ where $P\left(\bf{r}\right)$ is the polarization field due to the organic dipoles of the FE interacting with the electric field $E\left(\bf{r}\right)$ in the semiconductor from WE. In the work reported here we modified this model to consider the quadrupole exciton 'yellow' $1S$ level in $Cu_2O$, which is well known from many studies on bulk $Cu_2O$. This immediately puts our attention on a different interaction term, which is now $Q_{i,j}\frac{\partial}{\partial_i}E_{j,\bf{k}}$. Here the quadrupole field couples to a spatially varying (or $k$ dependent) electric field. This coupling is the basis of the present work. \footnote{In the point of view of Moskalenko, one can reconsider the coupling as dipole-dipole coupling, where the inorganic dipole moment is due to the 'quadrupole ' transitions. We show that the corresponding reinterpolation of the Agranovich formula fails.} In this paper we examine the new hybrid exciton which occurs in a $Cu_2O$/organic heterostructure. Although the oscillator strength of the quadrupole transition is three orders of magnitude smaller than the corresponding dipole case we show here how this can be compensated by strong spatial dispersion of the transition.      
\par
In the following section (\ref{configuration}), generalizing the work \cite{BASSANI:1999}  on dipole-dipole hybrid excitons, formed in adjacent layers of organic-inorganic  hetero-structures, we will here introduce a modification. The system we will analyze in this paper makes use of dipole forbidden, quadrupole $1S$ exciton in $Cu_2O$ (WE) coupled to a suitable organic Frenkel exciton (FE). From the fact of strong spatial dispersion of the quadrupole transitions we anticipate strong wave vector and polarization dependence of the dispersion for the hybrid. Concrete results will be given below for realistic configuration and values of parameters for particular organic materials.
\par 
To the best of our knowledge, there are no experimental studies yet on confinement effects for the cuprous oxide exciton in a quantum well. Therefore in the third (\ref{quasi}) section we present theoretical examination of confinement effects which are essential due to the rather small Bohr radius of $1S$ exciton; central cell and field effects on the exciton dispersion.
\par
In the fourth (\ref{organic}) paragraph we discuss the choice of some proper organic materials to assure resonance between FE and WE in the quantum wells. We are going to show that layers of PS:CA:DCM2 \cite{MADIGAN:2003} give an excellent match to the quadrupole 'yellow' exciton of $Cu_2O$. We have chosen this organic compound due to three main factors: 
\begin{enumerate}
	\item {the extremely big oscillator strength of the organic Frenkel exciton formed on DCM2 molecules;}
	\item {unlike the conventional organic with short emission lifetime, the dynamic 'solid state solvation' mechanism discovered in such a compound would allow the hybrid to live through the phase of 'slow' solvation ($\approx 3 \: ns$) which is comparable to the lifetime of the quadrupole exciton;}
	\item {the ability to tune into the resonance the energy of the singlet FE simply by means of changing concentration of the CA.}
\end{enumerate}
\par
The strong local electric field induced by the organic Frenkel excitons penetrates into the cuprous oxide layer and results in an induced 'confined Stark effect'. Relative shift of the electron and hole give rise to induced polarization and reduced hybridization effect. This will be discussed in detail in the fifth (\ref{stark}) section.
\par  
Hybridization requires coupling between the two excitons, and we will estimate the coupling coefficient in the sixth (\ref{dispersion}) section. The coupling between FE and WE in the case of quadrupole active $1S$ WE is due to gradient of the field induced by the FE in the DCM2 organic. In the dipole-dipole hybrid exciton one utilizes big oscillator strength of the FE and big Bohr radius of the WE, then in case of quadrupole-dipole exciton one utilizes big oscillator strength of the FE and long life time of the WE along with enhanced spatial dispersion of the coupling parameter. 
\par
In the final section (\ref{conclusion}) we will discuss the specifics of the quadrupole-dipole hybridization dispersion and briefly propose possible applications of such a hybridization. We look forward to experimental tests of our results, by means of new types of high precision spectroscopy which were invented recently \cite{FROHLICH:2005}.  

\section{\label{configuration} The quadrupole-dipole hybrid exciton configuration}

\begin{figure}[htbp]
\includegraphics[width=8cm]{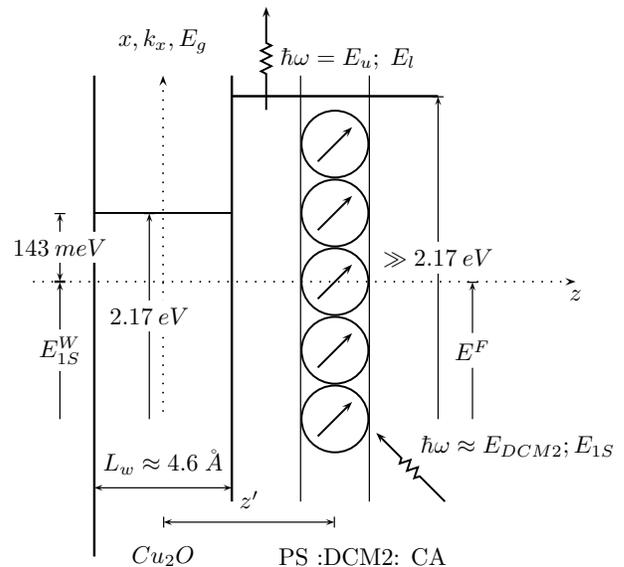}
	\caption{Schematic representation of a possible experimental set-up to produce quadrupole-dipole excitons. Here the inorganic $Cu_2O$ quantum well provides the Wannier-Mott $1S$ quadrupole exciton with the binding energy $143 \; meV$ (for details see section \ref{quasi}). Two pumping photons $\hbar \omega \approx E_{DCM2}, E_{1S}$ generate the hybrid signal from the upper $E_u$ and lower $E_l$ branches. The DCM2 part of the organic 'solid state solute' provides dipole allowed Frenkel exciton; the PS host prevents wave function overlapping between organic and inorganic excitons; CA under proper concentration allows tuning of the excitons into the resonance. Due to comparable lifetimes of both types of  exciton the system utilizes strong spacial dispersion of the quadrupole exciton and big oscillator strength of the organic.}
	\label{fig:1}
\end{figure}

Our proposed configuration of an experiment for obtaining the hybrid exciton is shown in the following diagram (see Fig.\ref{fig:1}).
In this simplified model a mono-layer of width $L_w\approx$ size of a unit cell $4.6 \; \AA$ quantum well of $Cu_2O$ (gap energy $E_g = 2.17 \; eV$ \cite{FROHLICH:2005}) is placed upon a thin film of the PS:CA:DCM2 organic (with the gap energy much bigger than that of the cuprous oxide). Obliquely incident (to assure $x$ component of the exciton wave vector) photons of close energies excite Wannier-Mott exciton in the cuprous oxide in resonance with the Frenkel exciton in DCM2. We consider DCM2 exciton as a 2D lattice of dipoles $d_x = 12D$  at discrete sites $\bf{n}$, placed at $z'\approx L_w/2$.
\par
 We treat the interaction of cuprous oxide quadrupole excitations only with DCM2 organic molecules in the quadruple approximation, since at the pumping laser frequency only dipole forbidden transitions are allowed. Due to the small concentration of the DCM2 molecules there is a 'buffer' of PS between $Cu_2O$ and DCM2 so that one may neglect the exciton wave function overlap and assume perfect 2D invariance of the system in direction transverse to growth. Further on we neglect kinetic energy of the Wannier exciton. Indeed, as will be shown in detail in the next paragraph, due to fluctuations of the inorganic quantum well (IQW) width, strong confinement of the exciton occurs \footnote{Let's consider smallest possible fluctuation of the IQW width, i.e. it changes from $L_w$ to $2L_w$ and decrease of the exciton energy is approximately 20 meV which is big comparing to the rather small kinetic energy of the exciton for it has rather big mass of free exciton} compared to the kinetic energy for small wave vectors which results in hopping motion of the exciton between sites of localization. 
 We seek the new quadrupole-dipole hybrid eigen-states for the upper ($u$) and lower ($l$) branches in the usual linear combination form:
\[
 \left| {u,{\mathbf{k}}} \right\rangle  = A_u\left| {F,{\mathbf{k}}} \right\rangle  + B_u \left| {W,{\mathbf{k}}} \right\rangle  \hfill ,\ 
  \left| {l,{\mathbf{k}}} \right\rangle  = A_l\left| {F,{\mathbf{k}}} \right\rangle  + B_l \left| {W,{\mathbf{k}}}\right\rangle  \hfill	
\]
where the weighting coefficients for small $\mathbf{k}$ are given by\cite{AGRANOVICH:1998}: $A_{u,l}^2=B_{u,l}^2=1/2$.
\par
After the usual diagonalization of the coupled WE/FE hamiltonian, $H=H_{WE}+H_{FE}+H_{int}$, the energies of the resulting upper and lower branches are given by:
\[
	E_{u,l}=E^{W(F)}\pm \Gamma_{\mathbf{k}}
\]
where we have introduced the quadrupole-dipole interaction Hamiltonian and corresponding off-diagonal hybridization parameter: $\Gamma \left( {\mathbf{k}} \right) \equiv \left| {\left\langle {W,{\mathbf{k}}} \right|\hat H_{int} \left| {F,{\mathbf{k}}} \right\rangle } \right|$.
\par
In the subsequent paragraphs we will derive expressions for the energies of the exciton and necessary conditions for the resonance hybridization. Also as a main result of the article we will derive an analytical expression for the hybridization parameter and dispersion. This will be followed by quantitative and qualitative comparison of systems with dipole-dipole hybridization \cite{AGRANOVICH:1998}. 

\section{\label{quasi} Quasi 2-D exciton in $Cu_2O$.}

As far as we are aware there is no experimental data or theoretical description for the film cuprous oxide systems; contrary to the extensive literature for the bulk case. In case of the small size $Cu_2O$ quadrupole exciton one can not consider strong confinement effect, which requires the width of the IQW much less than the Bohr radius of the exciton $a_b$. Even if modern epitaxy methods allows one to get a molecular mono-layer for the quantum well thickness, one would have
${{L_w } \mathord{\left/
 {\vphantom {{L_w } {a_b  \approx {1 \mathord{\left/
 {\vphantom {1 2}} \right.
 \kern-\nulldelimiterspace} 2}}}} \right.
 \kern-\nulldelimiterspace} {a_b  \approx {1 \mathord{\left/
 {\vphantom {1 2}} \right.
 \kern-\nulldelimiterspace} 2}}}$
 which is not enough for pure 2D consideration. Two and more mono layers in one quantum well will give the case of weak confinement and result in much weaker coupling (see below). Thus we start from the well described case of bulk "yellow" excitons and then estimate the main properties in case of strong confinement theoretically.
\par  
$Cu_2O$ condenses in a cubic structure, where the copper ions form a face-centered sub lattice, while the oxygen ions form a body-centered sub lattice. The arrangement of both sub lattices is such that a copper ion is found centered between two neighboring oxygen ions ($O_h$ symmetry) with the lattice constant $a \approx 4.26 \; \AA$. From the lattice structure we now turn to the band structure of this crystal \cite{ELLIOT:1961}. There is a direct band gap, where the valence band is formed by the Cu $3d$ orbitals and  the conduction band arises from Cu $4s$ and (possible) oxygen orbitals. When considering the cubic crystal field, the five $3d$ states of the valence band split further into three states of type $^3\Gamma _5^ +  $ and a twofold $^2\Gamma _3^ +  $ level. Taking also spin-orbit interaction into account, the state splits further into a twofold level $^2\Gamma _7^ +  $ and a fourfold degenerate $^4\Gamma _8^ +  $ level. The exciton representation is obtained from the direct product of electron and hole $\Gamma _{ex}  = \Gamma _{envelop}  \otimes \Gamma _e  \otimes \Gamma _h $. For S-excitons : $^1\Gamma _1^+ \otimes ^2\Gamma _7^+ \otimes \Gamma _6^+ = ^3\Gamma _5^{+} + ^1\Gamma _2^+  $
The threefold degenerate $^3\Gamma _5^ +  $ state and single $^1\Gamma _2^ +  $ state are termed ortho-exciton and para-exciton respectively. The Para-exciton is optically forbidden in bulk. The Ortho-exciton is dipole $\left( {^3\Gamma _4^ -  } \right)$ forbidden from the ground state of symmetry $\left( {^1\Gamma _1^ +  } \right)$, because $\left\langle {^1\Gamma _1^ +  } \right|\!^3\Gamma _4^ -  \left| {^3\Gamma _5^ +  } \right\rangle  = 0$ and couples to the light in lowest order via quadruple interaction of symmetry $\left( {^3\Gamma _5^ +  } \right)$, $\left\langle {^1\Gamma _1^ +  } \right|\!^3\Gamma _5^ +  \left| {\!^3\Gamma _5^ +  } \right\rangle  \ne 0$.
\par
Unlike the dipole operator, the quadrupole operator depends on the direction of the light wave vector ${\mathbf{k}}$ relative to the lattice and the radiation polarization vector  ${\mathbf{e}}$. Because of the ${\mathbf{k}}$ dependence the transition is anisotropic even in a cubic crystal. The amplitudes of the ortho-exciton quadruple transitions are given by the symmetric vector product of ${\mathbf{k}}$ and ${\mathbf{e}}$ : $ \sim {\mathbf{e}}_i {\mathbf{k}}_j  + {\mathbf{e}}_j {\mathbf{k}}_i ,{\text{     }}i \ne j$, see for example \cite{ELLIOT:1961}. The three components correspond to the Cartesian representations:  $^3\Gamma _{5xz}^ +  ,^3\Gamma _{5yz}^ +  ,^3\Gamma _{5xy}^ +  $.
\par
The measured \cite{GROSS:1952} oscillator strength of the quadrupole transition in bulk $Cu_2O$ is low $f \simeq 3.9 \cdot 10^{ - 9} $, which is about four orders of magnitude smaller than the value found for the dipole transitions of the P-excitons of the yellow series. Even though the coupling to the light is extremely weak, it can not be disregarded. The binding energy is about $153\:meV$ and Bohr radius of the exciton is given by $a_b  = \frac{{2\pi \varepsilon _0 }}
{{E_b }} \approx 7\mathop {\rm A}\limits^0 $.
\par
Because of the unidirectional confinement in the IQW, the exciton is discretized in this direction ($z$-direction). And so the symmetry group $O_h$ is reduced to $D_{4h}$. As a result the three fold degenerate ortho-exciton $^3\Gamma_5^+$ is split into two fold degenerate $^2\Gamma_5^+$ and non degenerate  $^1\Gamma_4^+$ ortho-exciton levels, which we are going to refer to as "heavy" and "light hole exciton" in analogy with the well known case of dipole allowed exciton. Another remarkable result of the confinement is that the para-exciton $^1\Gamma_2^+$ changes its symmetry to $^1\Gamma_3^+$. And due to the fact that the quadrupole operator $^3\Gamma_5^+$ also reduces its symmetry to $^2\Gamma_5^+$ and $^1\Gamma_3^+$ the para-exciton   is no longer forbidden  in the IQW. Note that roughness of the interface due to the small width of the IQW leads to further reduction of the symmetry to $D_{2h}$. The details of the selection rules and dependence upon polarization can be found in the Table I, II.  
\par
Now let us consider the effect of confinement on the energy spectrum of the quadrupole active exciton. We study only the 
ortho-exciton (the generalization to the case of para-exciton is trivial) and neglect spin-dependent exchange interaction and  related spin-orbit effects.
The energy of the $1S$ quadrupole exciton confined in a IQW can be written as: 
\[
	 H_{WE} =H_e \left( {z_e } \right)+H_h \left( {z_h } 
\right)+\frac{p^2}{2\mu }+\frac{P^2}{2M}-\frac{e^2}{\varepsilon 
\sqrt {\rho ^2+\left( {z_e -z_h } \right)^2} }
\]
where $\rho$ is the relative electron $e$ - hole $h$ position, $p$ is the relative momentum, $P$ is center of mass quasi momentum which is to be considered in $x,y-$ plane. $\mu$ and $M$ are reduced mass and total mass of the exciton respectively.
\par
\begin{figure}[htbp]
	\centering
		\includegraphics{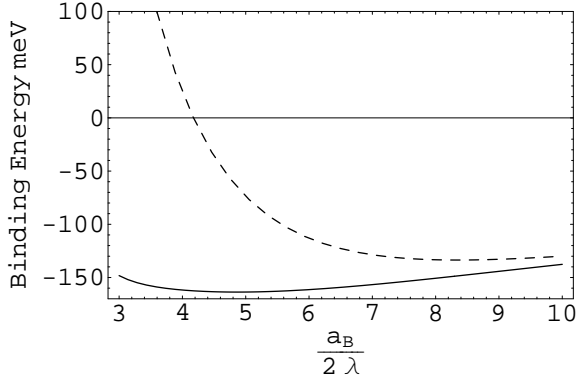}
	\caption{The solid line shows the modified energy of the $1S$ exciton as a function of the Bohr radius ($\AA$) and variational parameter $\lambda$ due to the confinement effect; the dotted line shows the same function with the central cell correction not taken into account.}
	\label{fig:2}
\end{figure}

The quadrupole exciton confined in the infinite IQW has a smaller Bohr radius then in case of bulk and as a result one must take into account so called "central cell" corrections (CCC) \cite{KAVOULAKIS:1997} in determining the dispersion (See Fig.\ref{fig:2}).

 Aside from the non-parabolicity of the bands \cite{MOSKALENKO:1959}, the electron-hole interaction in this case is the bare Coulomb interaction modified by the ${\mathbf{k}}$ dependent dielectric function. This is mainly due to virtual LO-phonon assisted valence electron transition into the highest conduction bands ( $^4\Gamma_8^-$ split by the confinement). Here we have to assume that the energy of the LO-phonon modes exceeds 87 meV \cite{TREBIN:1981}, when the dielectric constant drops from  $\varepsilon _0  = 7.5 \mp 0.2$  to $\varepsilon _\infty   = 6.46$. In this case it can be shown \cite{KAVOULAKIS:1997} that for small values of the exciton wave number $k$, $\varepsilon (k) \approx \varepsilon _\infty   - 0.18\left( {ka } \right)^2 $.
\par 
In our model due to the comparable sizes of the exciton radius and IQW width one can consider the confinement as a small perturbation to the pure 2D exciton energy. To explicitly show the perturbation part it is convenient to introduce a small variational parameter $\lambda$ ranging from 1 for pure 2D to $1/2$ for pure bulk cases, so that the confined exciton energy can be re-written as:
\begin{equation}
\label{eq10}
H_{WE} =H_{WE}^0 + H_{WE}^1 + H_{WE}^2 
\end{equation}
Where we separate  the analytically solvable part: 
\begin{equation}
\label{eq:16}
 H_{WE}^0 =H_e \left( {z_e } \right)+H_h \left( {z_h } 
\right)+\frac{p^2}{2\mu }+\frac{P^2}{2M} 
-\frac{\lambda e^2}{\varepsilon \rho }
\end{equation}
and small perturbation parts are due to the weak confinement:
\begin{equation}
\label{eq:17} 
 H_{WE}^1 =\frac{\lambda e^2}{\varepsilon \rho }-\frac{e^2}{\varepsilon 
\sqrt {\rho ^2+\left( {z_e -z_h } \right)^2} } 
\end{equation}
and 
\begin{equation}
\label{eq:18}
	H_{WE}^2 = \frac{p^4a^2}{24\hbar ^2{\mu}'}-\frac{p^2P^2a^2}{4\hbar ^2{M}'}
\end{equation}.
These terms (\ref{eq:17},\ref{eq:18}) come from second order corrections in the tight-binding model and are due to non-parabolicity of the bands with $1/{\mu }'=C_e /m_e +C_h /m_h $ and ${M}'=m_e /C_e +m_h /C_h $. The second term in (\ref{eq:18}) couples the relative motion of the electron and hole with the motion of their center of mass and modifies the total exciton mass. The fourth term in (\ref{eq:16}) and second term in (\ref{eq:18}) yield the free exciton dispersion relation for the center of mass motion \cite{KAVOULAKIS:1997}:
\begin{equation}
\label{eq11}
E_{\rm {\bf k}} =\frac{\hbar ^2k^2}{2M}\left( {1-\frac{2 \lambda 
^2Ma^2}{{M}'a_b^2 }\frac{I_4 \left( {\frac{k_D a_b }{2\lambda }} 
\right)}{I_2 \left( {\frac{k_D a_b }{2\lambda }} \right)}} \right)
\end{equation}
Where $k_D$ is the Debye wave vector, i.e. the radius of a sphere with volume equal to that of the first Brillouin zone and $I_n \left( x \right)=\int\limits_0^x {\frac{y^ndy}{\left( {1+y^2} \right)^4}} $. 
Kouvalkis et. al.\cite{KAVOULAKIS:1997} estimated the order of magnitude of the constants $C_e,C_h$ using ${\rm {\bf k}}\cdot {\rm {\bf p}}$ perturbation theory. Mixing of bands with different parities modifies the bare electron and hole masses. 
In addition including the coupling to LO modes gives $C_e \approx C_h \approx1$.
\par
\begin{figure}[htbp]
	\centering
		\includegraphics[width=8cm]{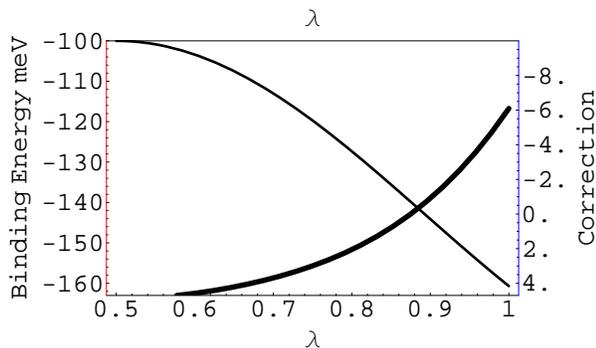}
	\caption{(Color online) For numerical estimation of the confined quadrupole Bohr radius as an approximation we took $a_b$ to be the bulk Bohr radius $5.1 \; \AA$. So the 'correction' equation (\ref{eq13}) becomes function of only one parameter $\lambda$ and so does the binding energy (\ref{eq12}).}
	\label{fig:3}
\end{figure}

 In our work we are going to use a simplified approach instead the ${\rm {\bf k}}\cdot {\rm {\bf p}}$ perturbation theory and estimate this correction from the fact that the exciton is becoming localized as the Bohr radius approaches the lattice constant: $\mathop {\lim }\limits_{a_b \to a} E_{\rm {\bf k}} =0$. This yields the following expression $C=2I_2 \left( \pi \right)/I_4 \left( \pi \right)$. The correction to the potential energy 
due to the first term in (\ref{eq:18}) is given by its expectation value in momentum space. 
\begin{equation*}
\begin{split}
-\frac{\hbar ^2a^2}{24{\mu }'}\left( {\sum\limits_{q<k_D } {\left| {\Psi 
_{\rm {\bf q}} } \right|^2q^4} } \right)\left( {\sum\limits_{q<k_D } {\left| 
{\Psi _{\rm {\bf q}} } \right|^2} } \right)^{-1}=\\
= - \frac{\hbar ^2a^2C}{24\mu 
}\frac{I_6 \left( {\frac{k_D a_b }{2\lambda }} \right)}{I_2 \left( 
{\frac{k_D a_b }{2\lambda }} \right)} 
\end{split}
\end{equation*}
Here the relative electron $e$ - hole $h$ motion wave function $\Psi _{\rm {\bf q}} =\frac{8\sqrt {\pi a_b^2 } }{\left( {1+\left( {q a_b } \right)^{^2}} \right)^2}$ is strongly peaked in momentum space. The last term in $H_{WE}^0$ describes the effective interaction between an electron and hole at momentum transfer $q$ with the same approximation for the ${\rm {\bf k}}$ dependent dielectric function as above: 
\[
V\left( q \right)=\frac{4\pi e^2}{q^2\varepsilon \left( q \right)}\approx 
\frac{4\pi e^2}{q^2\varepsilon _\infty }+\frac{4\pi e^2}{\varepsilon _\infty 
}a^2\cdot 0.18.
\]
If one truncates the $1S$ trial wave function outside the first Brillouin zone we find the correction due to the small Bohr radius of the exciton in the form:
\[
-\frac{0.36\lambda ^3e^2a^2}{\pi \varepsilon _\infty ^2 a_b^3 }\frac{\left[ 
{I_2 ^\prime \left( {\frac{k_D a_b }{2\lambda }} \right)} \right]^2}{I_2 
\left( {\frac{k_D a_b }{2\lambda }} \right)}
\]
In the above expression the contribution of LO phonons to the dielectric function has been employed, where the main contribution comes from virtual transitions between the higher $\Gamma _8^-$ conduction band and the highest $\Gamma _7^+$ valence band. Now treating $H_{WE}^1$ as perturbation in the lowest order (neglecting momentum dependence of the screening and approximating $H_e \left( {z_e } \right)+H_h \left( {z_h } \right)+\frac{P^2}{2M}$ by $\frac{\hbar ^2\lambda ^2}{\mu a_b^2 }$ ) one can find the total Wannier-Mott exciton energy in the IQW as the minimum of the total exciton energy with respect to the Bohr radius :
\begin{equation}
\begin{split}
\label{eq12}
E_{WE}^{total} \left( {\lambda ,a_b } \right)=\frac{\hbar ^2\lambda ^2}{\mu 
a_b^2 }-\frac{2\lambda e^2}{\varepsilon _\infty a_b }-\\
-\frac{\hbar ^2a^2C}{24\mu }\frac{I_6 \left( {\frac{k_D a_b }{2\lambda }} \right)}{I_2 
\left( {\frac{k_D a_b }{2\lambda }} \right)}-\frac{0.36\lambda ^3e^2a^2}{\pi 
\varepsilon _\infty ^2 a_b^3 }\frac{\left[ {I_2 ^\prime \left( {\frac{k_D 
a_b }{2\lambda }} \right)} \right]^2}{I_2 \left( {\frac{k_D a_b }{2\lambda 
}} \right)}
\end{split}
\end{equation}
To find the variational parameter $\lambda$ we use an additional restriction determined by requiring that the first 
order energy shift vanishes \cite{JAZIRI:1992}:
\begin{eqnarray}
\label{eq13}
\left\langle {\Psi _\lambda } \right|H_{WE}^1 \left| {\Psi _\lambda } 
\right\rangle =0 \\
\Psi _\lambda =\chi _e \left( {z} \right)\chi _h \left( {z} 
\right)\frac{\left( {2\lambda } \right)^{3/2}\exp (-\frac{2\lambda \rho 
}{a_b })}{\sqrt {\pi a_b^3 } }e^{-i{\rm {\bf kr}}_\parallel }
\end{eqnarray}
Where $\Psi_\lambda$ is the unperturbed eigenfunction of $H_{WE}^0 $. The envelope functions for the confined electron and hole are denoted as $\chi _e$ and $\chi _h$ correspondingly and for $k=0$ given by equation (\ref{ISE:9}). The numerical calculations (see fig.\ref{fig:2} and fig.\ref{fig:3}) show that for a  mono-layer of the cuprous oxide $\lambda \approx 0.881, \; a_b \approx a=4.6 \mathop {\rm A}\limits^0 $, i.e. in this case one can consider the $1S$ quadrupole exciton to be Frenkel like localized exciton with $k$ dependent oscillator strength.
\par
Without central cell corrections the value of the binding energy is $124 \; meV$ (minimum of the dashed curve on Fig.\ref{fig:2}). With CCC it lowers to $154 \; meV$ and it corresponds to the minimum on the lower solid curve of Fig.\ref{fig:2}. Now in the weak confinement of the IQW there is a restriction on the value of the parameter $\lambda$ given by equation (\ref{eq13}). So the standard binding energy (second term in the total energy (\ref{eq12})) slightly grows with the increasing parameter $\lambda$ \footnote{$\lambda = 1/2$ for the bulk, and $\lambda =1$ for the maximum of the confinement (strong confinement regime)}. But the CCC with increasing confinement $\lambda$ decreases. This results in binding energy for weak confinement bigger than bulk case without CCC but slightly smaller then for the bulk case with CCC. This unusual behavior is entirely attributed to the confinement dependent CCC.
\par
The confinement in the IQW increases the overlap of the electron and hole wave functions, which in turn increases the oscillator strength of the 'yellow' transition. We are only able to estimate the oscillator strength in case of quantum confinement. We consider a non radiative interface exciton, which refers to the states outside the photon cone, $k\ge \frac{\omega \sqrt \varepsilon }{\hbar c}$. The exciton propagating in the plane are trapped and accompanied by the light field which is evanescent in the direction perpendicular to the interface (if one considers formation of polariton modes), i.e., are invisible at macroscopic distances from the IQW. In the strictly two-dimensional limit, the oscillator strength of the lowest state scales as $\frac{f^{2D}}{S a_b }\sim \frac{\left( {\lambda_k } \right)^{-3}}{a_b^3 }$. The oscillator strength per unit volume scales as $\frac{f^{3D}}{V}\sim \frac{1}{a_b^3 }$. The maximum enhancement of the oscillator strength in IQW with respect to the bulk case 
is eight times and results in giving $f^{2D}\approx 8\cdot 3.7\times 10^{-9}$. In general, due to the interaction with the Frenkel exciton $f^{2D}$ will have weak dependence on the wave vector. Also the exciton resonance broadens due to imperfections of the IQW. In our case, the IQW is rather thin, which gives rise to interface roughness and thickness fluctuations. Therefore the exciton mode should show a line width of about $1\: \mu eV$ (the lifetime $\approx 1.7 \: ns$).
\section{\label{organic} PS:CA:DCM2 as an organic part of the hybrid}
As it was already discussed, for the best manifestation of the hybridization effect one has to be able to tune the energy of the Frenkel and Wannier-Mott excitons into resonance. Also the Frenkel exciton lifetime should be comparable to the lifetime of the 1S quadrupole exciton, otherwise the hybrid would not live long enough to utilize peculiar properties of the quadrupole part. So we decided to use in our model the Frenkel exciton formed in a recently reported amorphous organic thin film doped with the red laser dye: [2-methyl-6-2-(2,3,6,7-tetrahydro-1H, 5H - benzo[i,j] - quinolizin - 9 - yl) - ethenyl] - 4H - pyran - 4 - ylidene] propane dinitritle (DCM2)\cite{BULOVIC:1999},
which has an electric dipole moment $d_x^F\approx 11D$ in the ground state. To achieve a red spectral shift of DCM2 into the resonance with the 1S quadrupole exciton we propose to use a low DCM2 dopant concentration in a two component host consisting of polystyerine (PS) and the polar small molecule camorphic anhydride (CA). 
\par
Because DCM2 and CA (dipole moment $\approx 6D$) are highly polar molecules, and PS is a nearly non-polar (less then 1D), one can adjust the spectral shift by means of adjusting the relative concentration of the DCM2 and CA molecules. Increasing the DCM2 concentration one increases the strength of the local electric fields present in the film. In our case we will keep DCM2 concentration constant and low (0.05\%), to avoid overlaping of the organic and inorganic excitons, and limiting DCM2 aggregation effects. At the same time , the dielectric properties of the film are modified by changing the concentraton of CA, which has a large ground state dipole moment relative to its molecular weight and is opticaly inactive in the relevant region of the DCM2 photoluminiscence. The dielectric permitivity increases with increasing CA concentration linearly:
	\begin{equation}
	\label{eq:14}
	\tilde\epsilon=2.44+0.13\left(CA\%\right)
  \end{equation}
In contrast, the index of refraction $n \approx 1.55$ of the film is nearly constant. 
The PS provides a transparent , non polar host matrix. Such a mechanism for the spectral shift was termed 'solid state solvation', as the solvation mechanism underlying 'solvatochromism' of organic molecules in liquids \cite{BALDO:2001}. 
\par
\begin{figure}[htbp]
\includegraphics{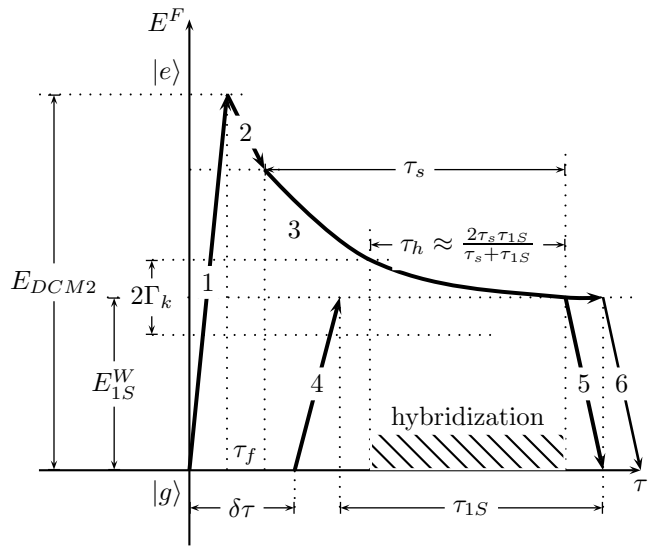}
	\caption{Schematic of dynamic tuning of Frenkel and $1S$ quadrupole Wannier excitons by means of 'solid state solvation' (SSS) red shift effect. Following photon absorption (1) in DCM2 there are distinct dynamic phases: (2) fast adjustment of the electronic configuration in DCM2 due to interaction with polar CA molecules within the time frame of pico seconds $\tau_f$; (3) slow self-adjustment of DCM2 and CA molecules $\tau_s \approx 3.3 \: ns$; (4) second 'yellow' photon delayed by $\delta\tau\approx\tau_s-\tau_{1S}$ is absorbed by $Cu_2O$ confined $1S$ quadrupole exciton with lifetime $\tau_{1S} \approx 1.7 \: ns$; hybridization occurs when detuning between this two types of exciton becomes smaller then the $k$ dependent coupling parameter, lifetime of the hybrid is defined by $\tau_s$ and $\tau_{1S}$; (5) phosphorescence due to the hybrid exciton recombination. (6) shows possible recombination of non-interacting excitons.}
	\label{fig:10}
\end{figure}

The theory of 'solvatochromism' relates the experimentally observed changes in emission  and absorption spectra of a solute (DCM2) to the dielectric permitivity of a solvent (PS:CA). An electronic transition on the solute (due to photon absorption) produces a corresponding change in the solute charge distribution, which causes the surrounding solvent molecules to respond to this new field  apart from the Frank-Condon (FC) shift due to solute nuclear reorganization (See Fig.\ref{fig:10} for details) in two ways:
\begin{enumerate}
	\item {through electronic cloud reorganization (polarizability) which is referred to as 'fast' solvation which occurs during $0.16 \; fs$ judging from the relaxation spectrum of the DCM2 (line width $\approx 0.25 \; eV$);}
	\item {through gross spatial movement due to physical translation and rotation ('slow' solvation). During this phase the radiative recombination from the FE is prohibited and allows the FE exciton to live $3.3 \; ns$. In our proposed sheme , the actual hybridization with 1S quadrupole exciton occurs during this time interval as it is comparable to the life time of the quantum confined quadrupole exciton, while the lifetime of the DCM2 in vacuum is determined by FC effect which is much faster;}
\end{enumerate}
Once the Frenkel exciton energy falls into the non zero coupling parameter vicinity with the quadrupole exciton (resonance condition) the hybridization occurs. The detailed dynamics of the hybrid is the subject of our future work. Presently we assume that the effective lifetime of the hybrid is $\frac{\tau_s \tau_{1S}}{\tau_s +\tau_{1S}}$ and requires two photons to populate both branches of the hybrid exciton. We performed calculations for the energy of the  Frenkel exciton under the 'continuum' approximation, in which the surrounding molecules of the solvent are replaced by a continues dielectric. The molecules of the solute are described by a spherical cavity of radius $a_F$ of the DCM2 molecules and corresponding charge distribution is reduced to the dominant dipole moment \cite{ONSAGER:1936}. Then the total emission energy including solvation effect can be written as:
\begin{equation}
\label{eq:15}
E=E_0+\Delta E_{solv}
\end{equation}      
where $E_0$ is the emission energy in vacuum including the FC shift; and $E_{solv} = -\frac{1}{a_F^3}(\mu_g-\mu_e)(\Lambda \mu_e+\Lambda_{op} \mu_g)$, here we introduced $\mu_e$ and $\mu_g$ to be ,respectively, solute excited and ground states dipoles, and
	\[
	\Lambda=\frac{2(\tilde\epsilon-1)}{2 \tilde\epsilon+1},\ \Lambda_{op}=\frac{2(n^2-1)}{2n^2+1}
  \]
associated with 'slow' and 'fast'	relaxation processes (described above) respectively. 
\par
The equations (\ref{eq:14},\ref{eq:15}) along with experimental fitting for $-\frac{1}{a_F^3}(\mu_g-\mu_e) \mu_e =0.57$ yields the concentration of the CA $ \approx 22 \% $ to correspond to the resonance with the confined $1S$ quadrupole energy $ 2.05\; eV $.
\par
There is one sensitive point we make in our work, namely necessity of the second delayed photon. One photon does allow hybridization once the energy of the FE exciton is close enough. Although only one branch of the hybrid is going to be populated. Because we approach the resonant energy from above, the upper branch is the one to be populated. The lower branch may be populated by multi acoustical phonon process. But first we do not know how long it takes to populate from the maximum of the upper branch to the minimum of the lower branch. For the temperature of $1.7 \; K$ it takes approximately $26$ of such phonon processes without Stark effect (see section \ref{stark}) or $7$ of them with the Stark effect. Even though such transitions are allowed it may take more time than life time of the hybrid. For example the lifetime of the $1S$ ortho-exciton itself is determined by the acoustical phonon assisted transformation to the para-exciton which involves change in energy $12 \; meV$. Which is comparable to $2 \Gamma_k$. 
\par
So the second photon is designed to assure both branches are populated regardless of the initial conditions. When both photons are in resonance with the WE exciton $\hbar \omega \approx E_{1S}$, the second photon provides conservation of the energy:
\[
\hbar \omega + \hbar \omega = \left(E_{1S} + \Gamma_k\right) + \left(E_{1S} - \Gamma_k\right)
\]  
\section{\label{stark} Induced Stark effect}
In this paragraph we investigate a new effect which plays a rather significant role in forming any hybrid exciton. Let us refer to it as an induced quantum confined Stark effect. The main idea is the following. There is an evanescent potential in $z$ - direction and corresponding ${\rm {\bf{k}}}$ - dependent electric field due to the Frenkel organic (DCM2) exciton (FE), penetrating into the inorganic quantum well (see Fig.\ref{fig:1}). This field induces some polarization of the bound electron-hole pairs forming the Wannier-Mott inorganic exciton (WE). This polarization leads to an effective screening of the electric field and results in reducing the coupling. 
\par
We consider the FE as polarization wave \cite{AGRANOVICH:1998} $\bf{P} \left( \bf{r} \right)$ confined to a perfectly 2D organic quantum well (OQW) placed at position ${z}'\approx \frac{L_w}{2}$. Neglecting the constant phase of the wave $e^{-i\frac{{\rm {\bf kL}}_w }{2}}$ the polarization per unit area $S$ is given as:
\begin{equation*}
{\rm {\bf P}}\left( {\rm {\bf r}} \right)= a_F {\rm {\bf 
d}}^F\frac{e^{-i{\rm {\bf kn}}_\parallel }}{\sqrt S }\delta \left( 
{z-\frac{L_w }{2}} \right)
\end{equation*}
Now let us we can move from discrete set of FE to a continuous distribution and take into account that
\begin{equation*}
\sum\limits_{\rm {\bf n}} {e^{-i{\rm 
{\bf kn}}_\parallel }} =\frac{1}{a_F^2 }\int {e^{-i{\rm {\bf kr}}_\parallel 
}d{\rm {\bf r}}_\parallel } 
\end{equation*} 
In this continuous approximation one gets the final result for the polarization created by the FE exciton propagation along the interface: 
\begin{equation*}
{\rm {\bf P}}\left( {\rm {\bf r}} \right)={\rm {\bf d}}^F\frac{e^{-i{\rm 
{\bf kr}}_\parallel }}{a_F \sqrt S }\delta \left( {z-\frac{L_w }{2}} 
\right)+c.c
\end{equation*}
The potential due to this polarization wave is given by \cite{AGRANOVICH:1998}:
\[
\varphi _{\rm {\bf k}} \left( {\rm {\bf r}} \right)=G\left( {{\rm {\bf 
r}},{\rm {\bf n}}_\parallel ,{\rm {\bf k}}} \right)\nabla {\rm {\bf 
P}}=-\nabla _i G\left( {{\rm {\bf r}},{\rm {\bf n}}_\parallel ,{\rm {\bf 
k}}} \right){\rm {\bf P}}_i 
\]
The standard Green's function for $z<{z}'$ is given by: 
\[
G\left( {z,{z}',{\rm {\bf 
k}}} \right)=\frac{4\pi }{\varepsilon \left( {\rm {\bf k}} \right)+\tilde 
{\varepsilon }}\frac{e^{k\left( {z-{z}'} \right)}}{k}
\]
It gives rise to the potential of the form: 
\begin{equation}
\label{ISE:4}
\begin{split}
\varphi _{\rm {\bf k}} \left( {\rm {\bf r}} 
\right)=\sum\limits_i {\left[ {k\delta _{i,z} +ik} \right]G\left( {{\rm {\bf 
k}},{z}',z} \right){\rm {\bf P}}_i \left( {\rm {\bf r}} \right)} = \\
= \varphi _{\rm {\bf k}} \left( {\rm {\bf r}} \right)={\rm {\bf 
d}}\frac{e^{-i{\rm {\bf kr}}_\parallel }}{a_F \sqrt S }\frac{4\pi 
}{\varepsilon +\tilde {\varepsilon }}e^{k\left( {z-{z}'} \right)} 
\end{split}
\end{equation} 
Here we utilized the fact that the complex conjugate part cancels out the imaginary part of the above expression. The electric field creates an additional polarization in the IQW, which we are going to estimate by its effect upon the coupling parameter. Effectively this polarization manifests itself in an envelope function equation in Ithe QW for the electron and hole. Following a standard procedure ( Bastard \cite{BASTARD}), the equations governing the electron (hole) envelope function in the IQW:
\begin{eqnarray*}
\left( {-\frac{\hbar ^2}{2m_e }\frac{d^2}{dz^2}-eF_{\rm {\bf k}} z 
}\right)\chi ^e_{\rm {\bf k}} \left( z \right)=\left( {E-E_c } \right)\chi^e_{\rm {\bf k}} \left( z \right) \\ 
\left( {-\frac{\hbar ^2}{2m_h }\frac{d^2}{dz^2}+eF_{\rm {\bf k}} z 
}\right)\chi ^h_{\rm {\bf k}} \left( z \right)=\left( {E_v -E} \right)\chi^h_{\rm {\bf k}} \left( z \right)  
\end{eqnarray*}
Where we used the long wave approximation for the potential namely taking the effective electric field in the form: $F_k \simeq k\varphi _{\rm {\bf k}}\left( {z={z}'} \right)$. The total energy shift due to the induced field is negligible for small difference in the electron and hole masses and small width of the IQW. The envelope functions are subject to zero boundary conditions at both sides of the IQW. Here one comes across some difficulties. Although this system has an exact solution the standard approach with Airy functions $Ai$ and $Bi$ will lead us to a rather complicated but exact result:
\begin{equation}
\label{ISE:8}
\begin{split}
\chi ^n_k \left( {z} \right)&=C_{1,n,k} Ai\left( {\frac{eF_k z-\zeta _{n,k}}{\zeta _{0,k} }} \right)+\\
 &+C_{2,n,k} Bi\left( {\frac{eF_k z-\zeta _{n,k}}{\zeta _{0,k} }} \right)
\end{split}
\end{equation}
In the last expression we introduce the following notation for the electron (hole) energy change due to the confinement and FE induced polarization:
\begin{eqnarray*}
\zeta _{0,k} =\left({\frac{\left( {eF_k \hbar } \right)^2}{2m}} \right)^{1/3} \\
\zeta _{n,k}=E_n -E_0 -e\varphi _{\rm {\bf k}} \left( 0 \right)
\end{eqnarray*}  
With $E_0 =0$ for the electron and $E_0 =-E_g $ for the hole (we also omitted indexes $e,h$ for simplicity). The normalization constants are connected as:
\[
C_{1,n,k} = \frac{Ai\left( {\frac{eF_k \frac{L_w 
}{2}-\zeta _{n,k} }{\zeta _{0,k} }} \right)}{Bi\left( {\frac{eF_k \frac{L_w 
}{2}-\zeta _{n,k} }{\zeta _{0,k} }} \right)}
\]
The energy levels can be found using boundary conditions as discrete solutions for:
\begin{equation}
\begin{split}
\label{ISE:7} 
Ai\left( {\frac{eF_k \frac{L_w }{2}-\zeta _{n,k} }{\zeta _{0,k} }} 
\right)Bi\left( {\frac{-\zeta _{n,k} -eF_k \frac{L_w }{2}}{\zeta _{0,k} }} 
\right)-\\
-Bi\left( {\frac{eF_k \frac{L_w }{2}-\zeta _{n,k} }{\zeta _{0,k} }} 
\right)Ai\left( {\frac{-\zeta _{n,k} -eF_k \frac{L_w }{2}}{\zeta _{0,k} }} 
\right)=0
\end{split}
\end{equation}
The constant is defined from normalization condition. For the hole one has to change $e\to -e$. Although (\ref{ISE:7}) and (\ref{ISE:8}) exhibit the exact solution, they are rather complicated for further analytical description of the hybrid exciton states. 
\par
There are some approximate ways to treat the problem of relative shift of the electron and hole wave functions in the induced field. First let us treat the electric field $F_k$ using perturbation theory. In this approach we can explicitly see the term due to induced polarization. This perturbation approach is applicable if the energy difference  between unperturbed ground and first exited state is much bigger then the perturbing potential at the average position of the particle $z=0$. For the case of a mono-layer of $Cu_2O$ it is applicable only for very small wave vector region but may be used as a correction in the dipole-dipole hybridization. In the first order there will be no change in total energy( $ \propto F_k^2$) but the wave functions will be changed to: 
\begin{equation}
\label{ISE:9}
\chi _k^e \left( z \right)=\sqrt {\frac{2}{L_w }} \cos \left( {\frac{\pi 
z}{L_w }} \right)-\frac{32 F_k L_w^3 em}{27 \pi ^2\hbar ^2}\sqrt {\frac{1}{L_w 
}} \sin \left( {\frac{2\pi z}{L_w }} \right)
\end{equation}
Because (\ref{ISE:9}) is still rather complicated, we will consider only the lowest energy levels transitions so it is possible to consider instead of the infinite IQW, an 'equivalent' parabolic profile defined through its lowest energy level as 
\[
\frac{1}{2}\hbar \omega _0^2 = \frac{\left( {\pi \hbar } \right)^2}{2mL_w }
\]
So we are able to consider ${\rm {\bf k}},L_w $ values in the region where the perturbation theory is not applicable. This problem has the exact solution with the same energy as was given by the perturbation theory for the lowest transition. The envelope function normalized inside the given IQW has a Gaussian form:
\begin{equation}
\label{ISE:10}
\chi _{\rm {\bf k}}^{e,h} \left( z \right)=\frac{\left( {2\pi } 
\right)^{1/4}}{\sqrt {erf\left( {\pi \mathord{\left/ {\vphantom {\pi {\sqrt 
2 }}} \right. \kern-\nulldelimiterspace} {\sqrt 2 }} \right)L_w } }\exp 
\left( {-\frac{\left( {z-z_{e,h} } \right)^2}{R_0^2 }} \right)
\end{equation}
In (\ref{ISE:10}) the dimension $R_0 $ of the parabolic IQW for the electrons and holes is taken to be the same:
\begin{equation*}
 R_0 =\sqrt {\frac{\hbar }{m_e \omega _0 }} =\frac{L_w }{\pi }
\end{equation*}
The shifted average electron and hole positions are given by:
\[  
 z_{e,h}=\pm \frac{eF_k^2 }{2m\omega _0^2 }=\pm \frac{eF_k^2 mL_w^4 }{2\pi ^4\hbar^2}
\]
In the next paragraph we are going to use these results for the envelope function to calculate the coupling parameter.

\section{\label{dispersion} The coupling parameter and dispersion}
The energy of interaction between the organic dipole and inorganic $1S$ quadrupolar excitons can be written as \cite{JACKSON:1975}:
\begin{equation*}
\begin{split}
\hat{H}_{int} =-\frac{1}{6}\sum\limits_\alpha {\sum\limits_\beta {\hat{Q}_{\alpha \beta } \frac{\partial }{\partial 
x_\alpha }} \hat{F}_\beta } = \\
= - \frac{1}{6} \hat {Q}_{\alpha ,\beta } \left| 0 \right \rangle D_{i,\alpha ,\beta } \left( 
{{\rm {\bf n}} - {\rm {\bf r}}_\parallel ,{z}',z} \right) \left\langle 0 
\right| \hat {d}_i^F 
\end{split} 
\end{equation*}
Where $\hat{Q}_{\alpha \beta }$ is a quadrupole transition operator; the electric field operator $\hat{F}_\beta$ is due to the Frenkel exciton as in (\ref{ISE:4}); $D_{i,\alpha ,\beta }$ is the Green's function in momentum space. The interaction parameter is given by the following transition matrix element:
\begin{equation*}
\begin{split}
\Gamma_{\bf k} = \left\langle {W,{\rm {\bf k}}} \right|H_{int} \left| {F,{\rm {\bf k}}} \right\rangle = \\
= \sum \limits_{\alpha ,\beta ,i} {\sum\limits_{\rm {\bf n}} 
{\int {dzd{\rm {\bf r}}_\parallel \left\langle {W,{\rm {\bf k}}} \right| {\hat{H}_{int}} \left| {F,{\rm {\bf k}}} \right\rangle }}}
\end{split} 
\end{equation*}
The quadrupole transition matrix element may be obtained by using the relation 
$
\frac{1}{S}\sum\limits_{\rm {\bf q}} {\varphi _{1S} \left( {\rm 
{\bf q}} \right)} =\varphi_{1S} \left( {{\rm {\bf r}}_e -{\rm {\bf r}}_h =0} 
\right)
$
between the wave function of the relative electron-hole motion of the $1S$ exciton $\varphi_{1S}$ normalized to the unit are $S$ and its momentum representation $\varphi_{1S,{\bf q}}$. Hence, the Fourier component of the quadrupole exciton transition from the ground state of the crystal \cite{MOSKALENKO:2002} is given as:
\begin{equation*}
\begin{split}
\left\langle {W,{\rm {\bf k}}} \right|\hat {Q}_{\alpha ,\beta } 
\left| 0 \right\rangle = \frac{e^{i{\rm {\bf kr}}_\parallel }}{\sqrt 
S }\sum\limits_{\rm {\bf q}} 
{\varphi _{1S} \left( {\rm {\bf q}} \right)Q_{\alpha,\beta } \left( {\rm {\bf 
q}} \right)} = \\
= \sqrt {\frac{2}{\pi }} \frac{\lambda Q_{\alpha ,\beta } }{a_b}\frac{e^{i{\rm {\bf kr}}_\parallel }}{\sqrt 
S }\chi ^e_k\left( z \right)\chi ^h_k\left( z \right)
\end{split}
\end{equation*}
Considering the weak dependence of the conduction $u_{c,q}$ and valence $u_{\upsilon,q}$ Bloch functions on the wave vector ${\bf q}$, the quadrupole moment of the inter-band transition may be written  as: 
\begin{equation*}
Q_{\alpha ,\beta } \approx Q_{\alpha ,\beta } \left( {\rm {\bf q=0}}\right)=\frac{1}{\upsilon _0 }\int\limits_{\upsilon _0 } {\mbox{d}{\rm {\bf s}}u_{c{\rm {\bf q}}}^\ast \left( {\rm {\bf s}} \right)x_\alpha x_\beta } 
u_{\upsilon {\rm {\bf q}}} \left( {\rm {\bf s}} \right)
\end{equation*}
with the integration taken over the unit cell of the cuprous oxide $\upsilon _0$.
The expectation value of the FE exciton dipole transition operator is given as:  
\begin{equation*}
\left\langle 0 
\right|\hat {d}_i^F \left( {\rm {\bf n}} \right)\left| {F,{\rm {\bf k}}} 
\right\rangle =\sqrt {\frac{\pi }{S}} a_F d_i^F e^{-i{\rm {\bf kn}}}
\end{equation*} 
The Green's function in momentum space for the given geometry is given by
\begin{equation*}
\begin{split}
D_{i,\alpha ,\beta } \left( {{\rm {\bf k}},{z}',z} \right)=\frac{4\pi 
}{\varepsilon +\tilde {\varepsilon }}k^2 e^{k\left( {z-{z}'} \right)} \times \\
\times \left({\frac{ik_i }{k}+\delta _{i,z} } \right)\left( {\frac{ik_\beta }{k}+\delta 
_{\beta ,z} } \right)\left( {\frac{ik_\alpha }{k}+\delta _{\alpha ,z} }\right) 
\end{split}
\end{equation*}
For the selection rules based upon the specific geometry of the problem and the corresponding Green's function see Table III.
Without loss of generality we may take the wave vector along the $x$ - axis. From the Green's function; polarization selection rules \cite{LIU:2005}; our specific type of interaction energy and the requirement of having real eigenvalues one can conclude that the only 
possible direction of the wave vector and dipole orientation is $ d_x^F $ and $Q_{x,z}$ respectively. 
\par
We note here that there is another possible approach to the problem of resonant interaction 
between these two types of exciton. If one modifies results of Moskalenko \cite{MOSKALENKO:2002} for our case and 
introduce an effective polarization due to quadrupole transitions via a 
${\rm {\bf k}}$ dependent inter-band dipole element 
\begin{equation}
\label{eq:20}
{\rm {\bf d}}_{ij,{\rm {\bf k}}} \sim \left| {Q_{ij} } \right|\left( {{\rm 
{\bf e}}_i k_j +{\rm {\bf e}}_j k_i } \right)
\end{equation}
Then one uses the  corresponding interband polarization, this would coincide with a case of 'dipole-dipole' transitions, when the electric field created by the set of WE dipoles will interact with the dipoles in the organic. In this way one could directly use the results of Agranovich \cite{AGRANOVICH:1998}. Due to an additional requirement for the Hamiltonian to be hermitian, it would give us 
the following result for the coupling constant:
\[
\Gamma _{i\ne j} \left( {\rm {\bf k}} \right)\sim Re\left( {{\rm {\bf 
d}}_{ij,{\rm {\bf k}}}^W D_{j\alpha } {\rm {\bf d}}_\alpha ^F } 
\right)\sim\left| {Q_{x,j} } \right|d_x^F k_x {\rm {\bf e}}_j {\rm {\bf e}}_x 
\]
Here we used the fact that the wave vector has only an $x$ component. And so 
this coupling parameter vanishes in this approximation as $j\ne x$. For that 
reason in this article we study interaction of the cuprous oxide quadrupole 
with the gradient of the electric field created by organic Frenkel exciton 
placed near the interface between the organic $PS:CA$ and inorganic $Cu_2O$.
\par
Calculating the integral of the coupling parameter and summing over the dipoles for the case when a perturbation approach is 
applicable yields: 
\[\Gamma \left( {\rm {\bf k}} 
\right)=\Gamma _1 \left( {\rm {\bf k}} \right)+\Gamma _2 \left( 
{\rm {\bf k}} \right)
\]
Here $\Gamma _1 \left( {\rm {\bf k}} \right)$- corresponds to the 
unperturbed coupling parameter, and the last term is due to the perturbation.
\begin{equation}
\label{eq:21}
\Gamma _1 \left( {\rm {\bf k}} \right)=\frac{8\sqrt {2\pi } }{6 \left( 
{\varepsilon \left( k \right)+\tilde {\varepsilon }} \right)L_w 
}\frac{ke^{-k{z}'}sinh\left( {\frac{L_w k}{2}} \right)}{\left( {1+\left( 
{\frac{kL_w }{2\pi }} \right)^2} \right)}\frac{Q_{xz} d_z^F }{a_F 
a_b L_w }
\end{equation}
To increase this coupling term one may consider the organic host PS:CA with DCM2 organized in a multi-layered structure.
Generalizing the same approach to this case of multi-layered organic with distance between the layers to be equal to 
twice the radius of the Frenkel exciton (compact composition) it can be 
shown that one has to multiply (\ref{eq:21}) by $\left\{ {\frac{2\cdot e^{2ka_F 
}}{e^{2ka_F }-1}} \right\}$. The factor of two in the last expression indicates 
that the organic is placed on both sides of our IQW and one must 
consider the symmetrical Green's function in the interaction. 
\par
Another term in the coupling parameter is due to induced Stark effect
\footnote{There's no induced Stark effect in multi-layered case due to the corresponding symmetry of the organic}
and has the form:
\[
\Gamma _2 \left( {\rm {\bf k}} \right)\sim -\frac{512}{27}\frac{L_w^4 
kF_{\rm {\bf k}} \cdot C\cdot M}{\pi ^2\hbar ^2}\frac{e\cdot \cosh \left( 
{\frac{kL_w }{2}} \right)}{k^4L_w^4 +10\pi ^2k^2L_w^2 +9\pi ^4}
\]
Where $F_{\rm {\bf k}} $ - is electric field induced by the organic layer of 
dipoles and the terms proportional to $L_w^6 $ have been omitted.
\par 
For the region of wave vector and width of the IQW where one can not use 
perturbation theory  another form of the coupling parameter can 
be introduced. In this case it is convenient to use the effective parabolic 
potential and the envelope functions in the form (\ref{ISE:10}), and the 
coupling parameter has the form:
\begin{equation}
\label{eq:23}
\begin{array}{l}
 \Gamma \left( k \right)\sim k^2e^{-k{z}'}\frac{\sqrt 2 }{4}\exp \left( 
{-\frac{16z_e^2 -R_0^4 k^2}{8R_0^2 }} \right)\times \\ 
 \times\left( {erf\left( {\frac{\sqrt 2 }{4}\left( {2\pi +kR_0 } \right)} 
\right)+erf\left( {\frac{\sqrt 2 }{4}\left( {2\pi -kR_0 } \right)} \right)} 
\right) \\ 
 \end{array}
\end{equation}
Here for simplicity we put $z_e \approx -z_h $, or in another words $m_e \approx m_h $.
For numerical estimation of the quadrupole matrix elements we 
used the well known \cite{MOSKALENKO:2002} result for corresponding oscillator strength:
\[
f_{xz,{\rm {\bf k}}_0 } =\frac{4\pi m_0 E_g }{3e^2\hbar ^2}\left( {\frac{a_b 
}{2\lambda a}} \right)^3{\rm {\bf e}}_z \cdot {\rm {\bf k}}_{0,x} \cdot Q_{x,z} 
\]
Where $\lambda $ represents affect of quantum confinement, and $\left| {{\rm 
{\bf k}}_0 } \right|=2.62\cdot 10^5cm^{-1}$.
\begin{figure}[htbp]
	\centering
		\includegraphics[width=8cm]{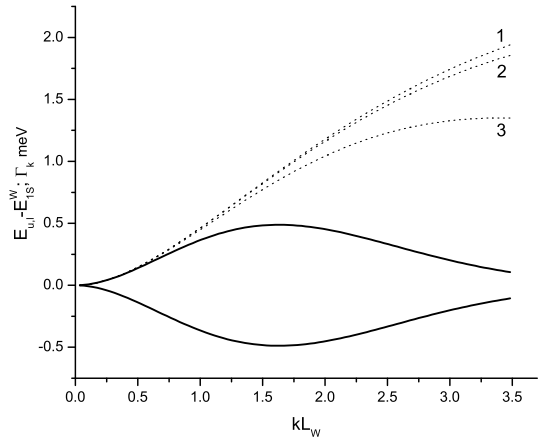}
	\caption{The solid lines represent upper and lower branches of the quadrupole-dipole hybrid dispersion when the coupling is calculated in the parabolic approximation and the induced Stark effect is taken into account; the dash lines correspond to the coupling parameter for different approximations: (1) - infinite IQW and the induced Stark effect is neglected, (2) - parabolic approximation, (3) - infinite IQW with the induced Stark effect treated as perturbation}
	\label{fig:4}
\end{figure}
\par
The dispersion of the hybrid state in case of negligibly small detuning of 1S exciton with the DCM2 transitions (resonance) is presented on Fig.\ref{fig:4}. The first (1) dashed lines correspond to the coupling parameter when the Stark effect is neglected. So it corresponds to to the upper part of the dispersion in this approximation. To show the effect of the 'parabolic' approximation on the coupling we drew the second dashed line (2). And the third dashed line is when the Stark effect is small enough to be taken as a perturbation without applying the 'parabolic' approximation.  
\par
Note that we consider the quadrupole-dipole interaction only between $Cu_2O$ and DCM2 excitons and neglect the coupling with PS:CA host transitions, as it has much smaller oscillator strength of the corresponding inter-band dipole moments for the transitions are off resonance with the inorganic. Neglecting the Stark effect from this numerical estimation it is easy to see that for the values of $k R_0<<1$ the expressions (\ref{eq:23}) and (\ref{eq:21}) are equivalent. 
\section{\label{conclusion} results and discussion}
Although the oscillator strength of the $1S$ quadrupole exciton is three orders of magnitude smaller than for the nearest dipole allowed exciton, strong spatial dispersion \footnote{factor of $k$ in equation (\ref{eq:21}) resulted from the quadrupole nature of the $1S$ transition in the cuprous oxide} makes the maximum of the coupling parameter comparable with those for the dipole-dipole case.
\par
The $1S$ quadrupole exciton has a rather big mass of the ($\approx 3 m_0$) and so is comparable to the mass of the Frenkel exciton ($\geq 5 m_0$). Also one has to take into account a significant effect of IQW width fluctuation. Indeed,  the effect of the width fluctuation is energy change between confined and the bulk cases, which corresponds to energy drop of $E_{1S,binding}\left(\lambda=1\right)-E_{1S,binding}\left(\lambda=0.881\right)=17 \; meV$ (See Fig.\ref{fig:3}). Hence , due to both of this effects we neglected the kinetic energy of the confined quadrupole $1S$ exciton (the motion occurs in Frenkel like way, by hopping between sites of localization). Thus the quadrupole-dipole hybridization effect is more pronounced than the dipole-dipole hybridization because it is not masked by kinetic energy of the exciton.
\par
Also the hybridization effect is not masked by the large radiative decay rate of the Frenkel exciton. Instead of the classical spontaneous radiative rate of the organic  we use the 'solid state solvation' process to obtain a different relaxation mechanism. Namely, one has the dynamical red shift of the DCM2 Frankel exciton during the 'slow' phase of the solvation process, when the energy of the photon excited DCM2 molecules is partially transfered to the polar CA molecules by non-radiative dipole-dipole e.m. interaction followed by immediate (compared to the $\tau_S$) radiative decay. To compare the radiative decay rate (life time) with the hybridization parameter we simplify the dynamics to the following statement. The Frenkel exciton has fixed resonant energy with the $1S$ quadrupole Wannier exciton but has effective lifetime equal to $\tau_S$ \footnote{Much better results for the decay rate of the hybrid may be obtained from a solution of the kinetic equations governing the dynamics of solvation and hybridization. This is a subject for our future work and will be reported elseware}. So if one omits all the possible non-radiative channels of decoherence except the radiative decay than the radiative decay rate from the hybrid to the ground state is calculated using Fermi's golden rule:
\begin{equation*}
\begin{split}
 \hbar \gamma _{hybrid} \left( {u,l;k} \right) = 2\pi \left| {\left\langle g \right|H_{{\mathop{\rm int}} } \left| {u,l;k} \right\rangle } \right|^2 \delta \left( {\hbar \omega _k  - E_{u,l} } \right) =  \\ 
  = \frac{{A_{l,u}^2 \hbar }}{{\tau _s }} + \frac{{B_{l,u}^2 \hbar}}{{\tau _{1S} }} = \frac{\hbar}{2}\frac{{\tau _s  + \tau _{1S} }}{{\tau _s \tau _{1S} }} \approx 0.29 \: \mu eV
 \end{split}
\end{equation*}
From this estimation the ratio of the hybrid radiative rate to the maximum of the coupling parameter is $0.0006$ which is two orders of magnitude smaller then predicted \cite{ENGELMAN:1998} value $0.09$ for the dipole-dipole hybrid in a quantum dot \footnote{We have not derived the dependence of the damping on the confinement and Bohr radius of the quadrupole-dipole hybrid. So we used the work of Engelman \cite{ENGELMAN:1998} on the fully confined quantum dot hybrid. And also work of V. Agranovich \cite{AGRANOVICH:1998} on the less confined quantum wire dipole-dipole confinement. This complies with our strategy to make comparative analysis between quadrupole-dipole and dipole-dipole hybrid cases}. Indeed it was shown in case of fully localized quantum dot dipole-dipole hybridization that the ratio of the hybrid exciton damping parameter to the coupling is proportional to $\frac{i \hbar \gamma_{hybrid}}{\Gamma_k} \propto \sqrt{a_b^3}$ ($\sqrt{a_b}$ - in case of quantum wires \cite{AGRANOVICH:1998}) so quadrupole-dipole hybridization takes advantage of the small radius of the $1S$ quadrupole. Also the fact of comparable life time of the organic and $1S$ quadrupole exciton significantly influences the ratio.
\par  
A noteworthy feature of the resulting upper (u) and lower (l) branch dispersion in Fig. \ref{fig:4} is that the well pronounced minimum on the lower branch can be populated by pump-probe experiment and now it may be possible to have finite critical temperature in case of quasi 2D excitons, due to the fact of non parabolic dispersion of the lower branch of the hybrid. This can provide a good basis for searching for BEC in such a hybrid. A new type of an interface polariton may also be expected.
\par
Note that we mentioned the possible BEC as a possibility only. Judging from the fact of the big saturation density of the hybrid and minimum of the energy on the lower branch. this minimum would make the hybrid exciton strongly localized and many dissipative processes (such as Auger heating \cite{KAVOULAKIS:2001}) are going to be suppressed. Also as we mentioned the non-parabolic dispersion allows the system to have a finite condensation temperature contrary to the case of $2D$ excitons with parabolic dispersion. But the main question about the life time of the hybrid is still a subject of our current research. So we do not make a concrete statement about the BEC of the hybrid but rather make an educated guess on that.
\begin{acknowledgments}
We have benefited from discussion with Prof. Que Nguyen Huong. Partial support from PCS-CUNY is greatly acknowledged.  
\end{acknowledgments}

\bibliography{quadrupole_dipole_hybridization}        
\bibliographystyle{apsrev}
\appendix
\begin{table*}[htbp]
	\centering
\begin{tabular}{|c|c|c|c|c|c|}
\hline
\multicolumn{6}{|c|}{$O_h$}\\
\hline
Symmetry & & $\begin{array}{c}\text{Electric dipole} \\ \text{operator} \end{array}$ & \multicolumn{2}{|c|}{$\begin{array}{c}\text{Electric quadrupole} \\ \text{operator} \end{array}$} &   Basis \\ 
\cline{3-5}
 & & $^3\Gamma_4^-$ & $^3\Gamma_5^+$ & $^2\Gamma_3^+$ & \\
\hline
Paraexciton & $^1\Gamma_2^+$ & F & F & F & $(x^2-y^2)(y^2-z^2)z^2-x^2)$\\
 \hline
Orthoexciton & $^3\Gamma_5^+$ & F & A & F & $xy;yz;zx$ \\ 
\hline
Basis &  & $x;y;z$ & $xy;yz;zx$ & $\begin{array}{c}2z^2-x^2-y^2; \\\sqrt{3}(x^2-y^2) \end{array}$ &  \\
\hline
\end{tabular}	
		\caption{Selection rules for the excitons in bulk $Cu_2O$. F-forbidden; A-allowed transition}
	\label{tab:1}
\end{table*}

\begin{table*}[htbp]
	\centering
	
\begin{tabular}{|c|c|c|c|c|c|c|c|c|}
\hline
\multicolumn{9}{|c|}{$D_{4h}$}\\
\hline
Symmetry & & \multicolumn{2}{|c|}{ $\begin{array}{c}\text{Electric dipole} \\ \text{operator} \end{array}$} & \multicolumn{4}{|c|}{$\begin{array}{c}\text{Electric quadrupole} \\ \text{operator} \end{array}$} &   Basis \\ 
\cline{3-9}
 & & $^1\Gamma_2^-$ & $^2\Gamma_5^-$ & $^2\Gamma_5^+$ & $^1\Gamma_3^+$ & $^1\Gamma_4^+$ & $^1\Gamma_1^+$ & \\
\hline
Paraexciton & $^1\Gamma_3^+$ & F & F & F & F & F & A & $(x^2-y^2)$ \\ 
\hline
Orthoexciton & $^1\Gamma_4^+$ & F & F & F & F & A & F & $xy$ \\
\cline{2-9}
 & $^2\Gamma_5^+$ & F & F & A & F & F & F & $ yz;zx $ \\ 
\hline
Basis &  & $z;x$ & $z$ & $yz;zx$ & $x^2-y^2$ & $xy$ & $R$ &  \\
\hline
\end{tabular}	
		\caption{Selection rules for the exciton in a $Cu_2O$ quantum well. F-forbidden; A-allowed transition}
	\label{tab:2}
\end{table*}

\begin{table*}[htbp]
\begin{center}
\begin{tabular}{|c|c|c|c|}
\hline
& 
$Q_{xy} $& 
$Q_{xz} $& 
$Q_{yz} $ \\
\hline
${\rm {\bf k}}_{\left[ {1,0,0} \right]} $& 
$\begin{array}{l}
 D_{xxy} =0 \\ 
 D_{yxy} =0 \\ 
 D_{zxy} =0 \\ 
 \end{array}$& 
$\begin{array}{l}
 D_{xxz} =-\frac{4\pi }{\varepsilon +\tilde {\varepsilon }}k^2e^{k\left( {z-{z}'} \right)},{\rm {\bf d}}=\left( {d_x ,0,0} \right) \\ 
 D_{yxz} =0 \\ 
 D_{zxz} =i\frac{4\pi }{\varepsilon +\tilde {\varepsilon }}k^2e^{k\left( {z-{z}'} \right)},{\rm {\bf d}}=\left( {0,0,d_z } \right) \\ 
 \end{array}$& 
$\begin{array}{l}
 D_{xyz} =0 \\ 
 D_{yyz} =0 \\ 
 D_{zyz} =0 \\ 
 \end{array}$ \\
\hline
${\rm {\bf k}}_{\left[ {0,1,0} \right]} $& 
$\begin{array}{l}
 D_{xxy} =0 \\ 
 D_{yxy} =0 \\ 
 D_{zxy} =0 \\ 
 \end{array}$& 
$\begin{array}{l}
 D_{xxz} =0 \\ 
 D_{yxz} =0 \\ 
 D_{zxz} =0 \\ 
 \end{array}$& 
$\begin{array}{l}
 D_{xyz} =0 \\ 
 D_{yyz} =-\frac{4\pi }{\varepsilon +\tilde {\varepsilon }}k^2e^{k\left( {z-{z}'} \right)},{\rm {\bf d}}=\left( {0,d_y ,0} \right) \\ 
 D_{zyz} =i\frac{4\pi }{\varepsilon +\tilde {\varepsilon }}k^2e^{k\left( {z-{z}'} \right)},{\rm {\bf d}}=\left( {0,0,d_z } \right) \\ 
 \end{array}$ \\
\hline
${\rm {\bf k}}_{\left[ {0,0,1} \right]} $& 
$\begin{array}{l}
 D_{xxy} =0 \\ 
 D_{yxy} =0 \\ 
 D_{zxy} =0 \\ 
 \end{array}$& 
$\begin{array}{l}
 D_{xxz} =0 \\ 
 D_{yxz} =0 \\ 
 D_{zxz} =0 \\ 
 \end{array}$& 
$
\begin{array}{l}
 D_{xyz} =0 \\ 
 D_{yyz} =0 \\ 
 D_{zyz} =0 \\ 
 \end{array}$ \\
\hline
\end{tabular}
\label{tab:3}
\caption{Green's function selection rules}
\end{center}
\end{table*}

\end {document}